\documentclass[preprint,aps,amsmath,superscriptaddress,nofootinbib,tightenlines,showpacs]{revtex4}
\usepackage{cases}
\usepackage{psfrag}
\usepackage{mathrsfs}
\usepackage{amsmath}
\usepackage{amsfonts}
\usepackage{amssymb}
\usepackage{latexsym}
\usepackage{graphicx}
\usepackage{bm}
\usepackage{epsfig}

\def\p{\partial}
\def\nn{\nonumber}

\def\csi{c_{\mathrm{s}I}}
\def\csj{c_{\mathrm{s}J}}
\def\csk{c_{\mathrm{s}K}}
\def\scro{\mathcal{O}}
\def\vp{\varphi}
\def\csn{C_{\mathrm{s}n}}

\usepackage{color}

\psfrag{p}{$\phi_1$}
\psfrag{q}{$\phi_2$}
\psfrag{dp1}{$\delta\phi_1$}
\psfrag{dp2}{$\delta\phi_2$}
\psfrag{vp1}{$\varphi_1$}
\psfrag{vp2}{$\varphi_2$}
\psfrag{s}{$\frac{\mathcal{P}_\zeta}{\mathcal{P}^{(0)}_\zeta}$}
\psfrag{c}{$\frac{c_{\mathrm{s}1}}{c_{\mathrm{s}2}}$}
\psfrag{f}{$\mathcal{P}_\zeta$}
\psfrag{g}{$\mathcal{P}_{\vp_1}$}
\psfrag{h}{$\mathcal{P}_{\vp_2}$}
\begin{document}


\title{Dynamics of Cosmological Perturbations in Multi-Speed Inflation}
\author{Shi Pi}\email{spi@pku.edu.cn}
\author{Du Wang}\email{wangdu14008@pku.edu.cn}
\affiliation{Department of Physics and State Key Laboratory of
Nuclear Physics and Technology, Peking University, Beijing 100871,
China}
\date{\today\\ \vspace{1cm}}
\begin{abstract}
We study a multi-field inflationary theory with separable Lagrangian, which has different speed of sound for each field. We find that the fields always coupled at perturbative level through gravitational interaction. We show that if the coupling terms among the perturbation fields are weak enough, these fields can be treated as a combination of decoupled fields, which are similar to normal modes in coupled oscillation. By virtue of such fields, the curvature perturbation at the horizon-crossing can be calculated up to the leading order of slow variation parameters via $\delta \mathcal{N}$ formalism. Explicitly, we consider a model of multi-speed DBI inflation, and calculate the power spectrum in detail. The result depends on the ratio of different speeds of sound, and shows an apparent amplification when the ratio deviates from unity.
\end{abstract}

\pacs{98.80.Cq, 04.50.Kd}

\maketitle

\newpage



\section{Introduction}\label{sect-intro}
Inflationary cosmology has become the prevalent paradigm to
understand the early stage of our universe, with its advantages of
resolving the flatness, homogeneity and monopole problems
\cite{inflation_bible, Starob}, and predicting a approximately scale-invariant
primordial power spectrum consistent with current cosmological
observations \cite{Komatsu:2010fb} very well. However, a single
field inflation model often suffers from fine tuning problems on the
parameters of its potential, such as the mass and the coupling
constant.

In recent years, people has noticed that, when a number of scalar
fields are involved in the inflationary stage, they can relax many limits on the single scalar
inflation model \cite{Liddle:1998jc}. Usually, these fields are able
to work cooperatively to give an inflationary stage long enough,
even none of them can sustain inflation separately. Models of this
type have been considered later in Refs. \cite{Malik:1998gy,
Kanti:1999vt, Copeland:1999cs, Green:1999vv}. The main results showed
that both the e-folding number $\mathcal{N}_e$ and the curvature perturbation
$\mathcal{R}$ are approximately proportional to the number of the
scalars $N$. Later, the model of N-flation was proposed by
Dimopoulos {\it et al.} \cite{Dimopoulos:2005ac}, which showed that
a number of axions predicted by string theory can give rise to a
radiatively stable inflation. This model has explored the
possibility for an attractive embedding of multi-field inflation in
string theory.

Over the past several years, based on the recent developments in
string theory, there have been many studies on its applications to
the early universe in inflationary cosmology. An interesting
inflation model, which has a non-canonical kinetic term inspired by
string theory, was studied intensively in the literature. Due to a
non-canonical kinetic term, the propagation of field fluctuations in
this model is characterized by a sound speed parameter and the
perturbations get freezed not on Hubble radius, but on the sound
horizon instead. One specific realization of this type of models can
be described by a Dirac-Born-Infeld-like (DBI) action
\cite{Aharony:1999ti, Myers:1999ps}. Based on brane inflation
\cite{Dvali:1998pa}, the model with a single DBI field was
investigated in detail\cite{Silverstein:2003hf, Chen:2004gc,
Chen:2005ad}, which has explored a window of inflation models
without flat potentials. In this model, a warping factor was applied
to provide a speed limit which keeps the inflaton near the top of a
potential even if the potential is steep.

Motivated by the effective description of multiple D-brane dynamics in string theory \cite{Taylor:1999pr}, an interesting inflation model involving multiple sound speeds with each sound speed characterizing one field fluctuation was proposed \cite{Cai:2009hw, Cai:2008if,Christopherson:2008ry}.
The authors of Refs.\cite{Cai:2009hw, Cai:2008if} suggest this scenario can be realized by a number of general scalar fields with arbitrary kinetic forms, and these scalars have their own sound speeds respectively. Therefore, this model is dubbed as ``{\it Multi-Speed Inflation}" (MSI). In this model, the propagations of field fluctuations are individual, and the usual conceptions in multi-field inflation models might be not suitable in this scenario. For example, in a usual generalized N-flation model, the length scale for perturbations being freezed takes the unique sound horizon; however, in our model it corresponds to the maximum sound horizon. It is worth emphasizing that the model of MSI is different from the usual DBI N-flation in which only multiple moduli fields are involved in one DBI action \cite{Ward:2007sj, Easson:2007fz, Huang:2007hh, Ward:2007gs, Langlois:2008mn, Contaldi:2008hr, Langlois:2009ej}, but MSI is constructed by multiple Kessence type actions. For example, an explicit model of MSI is made of two DBI fields in Ref \cite{Cai:2009hw}, and its primordial perturbations including curvature and entropy modes and
their non-Gaussianities were considered.

In the current paper we extend the model into a much more generic
case by relaxing the form of the Lagrangian. We find that even there
are no coupling terms among the inflaton fields, their perturbation
modes are coupled due to the nonlinear gravitational interaction. We
also find that if these couplings are weak, the independent
components of the fields can be treated as a combination of normal
modes, by which the curvature perturbation at the horizon-crossing
can be calculated in detail up to leading order of slow variation
parameters via $\delta \mathcal{N}$ formalism. As a specific example, we
consider a model of multi-speed inflation involving two DBI fields.
We show that in the relativistic limit, the coupling between two
fields is mainly reflected by the damping terms in the perturbation
equations. The difference between the sound speeds of these two fields is able to generate a
considerable amplification on the amplitude of the primordial curvature
perturbation. This is greatly different from the usual analysis on
the inflationary model constructed by two canonical fields\cite{Wands:2002bn, Gordon:2000hv}.

The paper is organized as follows. In Section II, we study the
cosmological perturbation theory of the MSI model in a generic case,
then show that the inflaton fluctuations could be decoupled through
a redefinition of fields at leading order under the assumption of
weak couplings. In Section III, we focus on a model of MSI involving
two fields, and study the detailed field transformation matrix to
illustrate that the decoupling process of the field fluctuations is
reliable under the weak coupling approximation. Specifically, we
analyze a specific model constructed by two DBI fields and give its
curvature perturbation. Section IV presents a summary
and discussion.

In the paper we take the normalization $M_p^2=1/8\pi G=1$ and the
sign of metric is adopted as $(-,+,+,+)$ in the following.

\section{The MSI Model}

An inflation model constructed with a single Kessence was originally
proposed by \cite{ArmendarizPicon:1999rj} and later its perturbation
theory was developed in \cite{Garriga:1999vw}. In the literature this
type of model has been widely studied, and one of the most
significant features is that there is an effective sound speed
describing the propagation of the perturbations. In the model of
MSI, for each k-essence field there is one sound speed
correspondingly. Therefore, the field fluctuations in this model do
not propagate synchronously due to different sound speed parameters,
but get coupled because of gravitational interactions. In the
current paper, our main interests focus on the effects of multiple
sound speeds and mode-couplings at the level of linear perturbation.
Before studying the perturbations, we first take an investigation on
the background equations.

\subsection{The setup of background model}\label{subsect-bgd}
The action is the summation of Hilbert-Einstein action and the action of $N$ different fields minimally coupled with gravity\cite{Cai:2009hw}
\begin{equation}\label{action}
    S=\int d^4x\sqrt{-g}\left[\frac{1}{2}R+\sum_{I=1}^NP_I(X_I,\phi_I)\right],
\end{equation}
with
\begin{equation}\label{X}
    X_I=-\frac{1}{2}g^{\mu\nu}\nabla_\mu\phi_I\nabla_\nu\phi_I.
\end{equation}
which is the kinetic term of the inflaton field $\phi_I$. From now on the subscript of capital Latin letter $I,J,...$ always denote the different inflaton fields. In this model, the Klein-Gordorn equations in a homogeneous background are
\begin{equation}\label{KG equation}
    \ddot\phi_I+\left(3H+\frac{\dot{P_I},_{X_I}}{P_{I,X_I}}\right)\dot\phi_I-\frac{P_{I,I}}{P_{I,X_I}}=0.
\end{equation}
Here and thereafter, the subscript ``$,I$'' denotes the derivative with respect to the field $\phi_I$, and ``$,X_I$" denotes the derivative with respect to $X_I$, for simplicity. Assuming a flat Friedman-Robertson-Walker background, the Friedman equations are given by
\begin{eqnarray}\label{Friedmann}
  3H^2 &=& \rho=\sum_I(2X_IP_I,_{X_I}-P_I), \\
  \dot{H} &=&
  -\frac{1}{2}(\rho+\sum_I P_I)=-\sum_IX_IP_I,_{X_I},
\end{eqnarray}
where $H$ denotes the Hubble parameter $\dot{a}/a$ at a given time.

To get analytical results, it is helpful to define some parameters to
characterize the kinetic behavior of the fields. In single-field
canonical inflation, a set of slow-roll parameters are necessary. To
ensure their smallness is not only the requirement of inflation to
occur and endure for long enough, but also some simplification in
analytical calculation. In a generic case, just as in
\cite{Chen:2006nt}, we define a similar set of slow-variation parameters.
Recall that in the simplest single-field inflation with canonical kinetic energy and flat potential, one can define two slow-roll parameters
\begin{eqnarray}
  \epsilon &=& -\frac{\dot H}{H^2} \sim \frac{\dot\phi^2}{V(\phi)} \sim \left(\frac{V'(\phi)}{V(\phi)}\right)^2,\\
  \eta &=& \frac{\dot\epsilon}{H\epsilon} \sim
  \frac{\ddot\phi}{H\dot\phi} \sim \frac{V''(\phi)}{V(\phi)}.
\end{eqnarray}
Here
%
%
the $\sim$ symbol denotes the approximate equality which only holds in
slow-variant environment and neglect the possible numerical factor. The second equality defines the slow-roll
parameter via the kinetic side, whereas the third equality defines
it in the potential viewpoint. If one only deal with the parameters
under slow-roll condition, for instance in a model with canonical kinetic terms, these definitions are almost equal, and
assuming one set is small can immediately assure the smallness of
the other sets. This is just because the dynamic behavior simply
relies on the shape of potential. But in a generic case, for
instance in DBI or k-inflation model, the dependence may be much more
complex and these different definitions are no more equivalent.
Since they are all very important in parameterizing our final
result, we have to define two sets of slow-variation parameters from different viewpoints. In
general, we should replace the slow-variation parameters by some
slow-variation matrixes that can reflect the correlations between
different fields. Since we are dealing with a separable Lagrangian,
we only need different parameters for each field. This is equivalent to the case with a diagonal slow-variation matrix. As is mentioned
above, one way to trace the motion is to define the parameter by
each kinetic term of one specific field:
\begin{eqnarray}\label{slow-roll:epsilonI}
\epsilon_I&=&\frac{P_I,_{X_I}X_I}{H^2},\\\label{slow-roll:etaI}
\eta_I&=&\frac{\dot{\epsilon_I}}{H\epsilon}.
\end{eqnarray}
In DBI model the field may roll down the potential very fast, but the parameter $\epsilon_I$ defined here is still small and slowly variant.  The summation of such parameters
gives the original version of them in canonical
case:
\begin{eqnarray}\label{slow-roll:epsilon}
\epsilon&=&-\frac{\dot{H}}{H^2}=\sum_I\epsilon_I,\\\label{slow-roll:eta}
\eta&=&\frac{\dot\epsilon}{H\epsilon}=\sum_I\eta_I.
\end{eqnarray}

Note that the magnitude of $\epsilon_I$ and $\epsilon$ obeys $\epsilon_I/\epsilon\sim\scro(1/N)$ in general. To accomplish our calculation we need another version of slow-variation parameters defined in the potential aspect:
\begin{eqnarray}\label{slow-roll:eI}
e_I&=&-\frac{P_I,_I}{3H^2},\\\label{slow-roll:hI} h_I&=&\frac{\dot
e_I}{He_I},
\end{eqnarray}
This is an analogue of $\epsilon\sim (V'/V)^2$, and for future convenience we take a squared root. It is not necessarily small in an infrared (IR) DBI model, in which the value of the inflaton field may be very small.
So for convenience let us focus on the ultraviolet (UV) DBI model where $\phi$ moves from UV side of the potential to the IR side. We should remind us that $e_I^2$  has the same order with $\epsilon_I$ in a model with canonical kinetic energy, but is very close to 1 in multi-DBI model. We refer to it as a half-order slow variation parameter in the former case, and tried to preserve terms higher than $e_I^2$. Besides,
\begin{equation}\label{slow-roll:iI}
 \iota_I=-\frac{P_I,_{II}}{3H^2}.
\end{equation}
as another extension of $\eta\sim V''/V$ should be defined. According to the speed of sound in the single-field case, we can define the speed parameter of each field:
\begin{eqnarray}
\csi^2&=&\frac{P_I,_{X_I}}{\rho_I,_{X_I}}=\frac{P_I,_{X_I}}{2X_IP_I,_{X_IX_I}+P_I,_{X_I}}\\
s_I&=&\frac{\dot{\csi}}{H\csi}
\end{eqnarray}
By this definition, the sound speed is the actually the effective propagation speed of the perturbations, which is different from the adiabatic sound speed which is defined as $c_{\mathrm{s(ad)}}^{2}=\dot{P}/\dot{\rho}$ when the cosmic fluid is described by scalar fields\cite{Christopherson:2008ry}. Notice that in a multi-Dirac-Born-Infeld model we have an extra relationship $P_I,_{X_I}=1/c_{\mathrm{s}I}$ which will connect this parameter with $\epsilon_I$. In general, $\csi$ needs not to be the same for all the fields, which makes this model different from the ordinary multi-DBI model and may give us some new phenomenon that we are about to investigate. For an analysis of $c_{\mathrm{s}}$ as an inverse Lorentz vector in a single-field DBI inflation, see \cite{Chimento:2010un}.

After having defined such parameters, the main quantities involving $P_I$ can be expressed in terms of these parameters directly from the inverse relations above. For instance,
\begin{eqnarray}
  P_I,_{X_I} &=& \frac{H^2}{X_I}\epsilon_I, \\
  P_I,_{X_IX_I} &=& \frac{\epsilon_I}{2}\left(\frac{H}{X_I}\right)^2\left(\frac{1}{\csi^2}-1\right), \\
  P_I,_I &=& -3H^2e_I,\\
  P_I,_{II} &=& -3H^2\iota_I.
\end{eqnarray}
Besides the $P$'s, another important quantity is the time derivative of one field. By using the
Klein-Gordon equation (\ref{KG equation}) we have
\begin{equation}\label{phidot}
\frac{\dot\phi_I}{2H}=\frac{\epsilon_I}{e_I}\left(1-\frac{1}{3}\epsilon\frac{\eta_I}{\epsilon_I}
-\frac{2}{3}\epsilon\frac{e_I}{h_I}\right).
\end{equation}

\subsection{Generic perturbation analysis}

In this subsection we perform a generic analysis on the linear perturbations of the model introduced previously. Since we are working in the frame of a cosmological system, the metric perturbations ought to be included as well as the field fluctuations. We would like to expand the action to the second order by virtue of the Arnorwitt-Deser-Misner (ADM) formalism \cite{Arnowitt:1962hi}.

To start, the spacetime metric in the ADM formalism is written as,
\begin{equation}\label{ADM metric}
    ds^2=-N^2dt^2+h_{ij}(dx^i+N^idt)(dx^j+N^jdt),
\end{equation}
the action can reads
\begin{equation}\label{action}
    S=\frac{1}{2}\int
    dtd^3x\sqrt{h}N\left(\frac{1}{2}R^{(3)}+\sum_IP_I\right)+\frac{1}{2}\int
    dtd^3x\sqrt{h}N^{-1}\left(E_{ij}E^{ij}-E^2\right)
\end{equation}
with $E_{ij}=(1/2)(\dot{h}_{ij}-\nabla_iN_j-\nabla_jN_i)$. When expanding the action up to the second order, we
can decompose the fields and Lagrangian multipliers as
\begin{eqnarray}
  \phi_I(t,\mathbf{x}) &=& \phi_I(t)+\delta\phi_I(t,\mathbf{x}), \\
  N &=& 1+\alpha, \\
  N_i &=& \p_i\beta.
\end{eqnarray}
First, the primary Hamiltonian and the momentum constrains gives two relations
\begin{eqnarray}
  \alpha &=& \frac{1}{2H}\sum_IP_I,_{X_I}\dot\phi_I\delta\phi_I,
  \\\nn
  \p^2\beta&=&\frac{1}{2H}\sum_{IJ}\left\{-\frac{P_I,_{X_I}}{\csi^2}\dot\phi_I\dot{\delta\phi}_I
  +\left(P_I,_I-2X_IP_I,_{IX_I}\right)\delta\phi_I\right.\\\nn
   &+&\left.\frac{P_I,_{X_I}}{H}\left(\frac{X_IP_J,_{X_J}}{\csi^2}\dot\phi_J\delta\phi_J
   -3H^2\dot\phi_I\delta\phi_I\right)\right\}.
\end{eqnarray}
Then we can make the perturbative expansion of the action up to second order
by using the Lagrangian constraints:
\begin{eqnarray}\label{S_2}\nn
    S_2&=&\int
    dtd^3x\frac{a^3}{2}\left\{P_I,_{X_IX_I}v_I^2+P_I,_{X_I}\dot{\delta\phi_I}^2
    -\frac{1}{a^2}P_I,_{X_I}\p_i\delta\phi_I\p_i\delta\phi_I
    -\frac{3}{2}\dot\phi_I\dot\phi_JP_I,_{X_I}P_J,_{X_J}\delta\phi_I\delta\phi_J\right.\\\nn
    &+&\frac{1}{H}P_I,_{X_I}\dot\phi_I\delta\phi_I\left(P_J,_{X_J}v_J+P_J,_{J}\delta\phi_J\right)
    +P_I,_{II}\delta\phi_I^2+2P_I,_{IX_I}v_I\delta\phi_I\\
    &+&P_I,_{X_I}\left.\left[3\dot\phi_I^2\left(\frac{1}{4H^2}P_K,_{X_K}P_L,_{X_L}
    \dot\phi_K\dot\phi_L\delta\phi_K\delta\phi_L\right)
    -\frac{2}{H}\left(P_J,_{X_J}\dot\phi_J\delta\phi_J\right)
    \dot\phi_I\dot{\delta\phi}_I\right]\right\},
\end{eqnarray}
with
\begin{equation}\label{vI}
    v_I=\dot\phi_I\dot{\delta\phi}_I-
    \left(\frac{1}{2H}P_J,_{X_J}\dot\phi_I\delta\phi_I\right)\dot\phi_I^2
\end{equation}
as the perturbation of $X_I$ up to first order. This result is in consistency with the one derived in the MSI model \cite{Cai:2009hw} and see \cite{Arroja:2008yy, Gao:2008dt} for a general case. The Lagrangian (\ref{S_2}) contains a lot of coupled terms as $\delta\phi_I\dot{\delta\phi}_J$ and $\delta\phi_I\delta\phi_J$.
These terms imply that the perturbations of inflaton fields depend on others during their evolution, and the equations of motion would be difficult to be solved directly. In this note we aim at making the transformation of the inflaton perturbations as to obtain approximately decoupled equations of motion for new fields. In principle this is difficult, if not impossible to realize. We will see in which case can we do such transformations and see what will happen by such a method. This will require some constraints on the slow-variation parameters. We will see this in detail for a double-field toy model.

After utilizing the equations of motion for background and taking
some integration by parts, one can write the Lagrangian in a more
compact form:
\begin{eqnarray}\nn
  S_2 &=& \frac{1}{2}\int dtd^3xa^3\left\{(P_I,_{X_I}+P_I,_{X_IX_I}\dot\phi_I^2)\dot{\delta\phi}_I^2
  -\frac{1}{a^2}P_I,_{X_I}\p_i\delta\phi_I\p_i\delta\phi_I\right.
  \\\label{action:2nd order}
   &+&
   \left.\mathscr{N}_{IJ}\dot{\delta\phi}_I\delta\phi_J-\mathscr{M}_{IJ}
   \delta\phi_I\delta\phi_J\right\}
\end{eqnarray}
with $\mathscr{N}_{IJ}$ and $\mathscr{M}_{IJ}$ are
time-dependent damping and mass terms,
\begin{eqnarray}
  \mathscr{N}_{IJ} &=& 2P_I,_{IX_I}\dot\phi_I\delta_{IJ}-\frac{1}{H}P_I,_{X_IX_I}
  \dot\phi_I^3\dot\phi_J,
  \\\nn
  \mathscr{M}_{IJ} &=& -P_I,_{II}\delta_{IJ}+\frac{1}{H}P_I,_{IX_I}P_J,_{X_J}\dot\phi_I^2\dot\phi_J\\
  &-&\frac{1}{4H^2}P_K,_{X_KX_K}\dot\phi_K^4P_I,_{X_I}P_J,_{X_J}\dot\phi_I\dot\phi_J
  -\frac{1}{a^3}\left(\frac{a^3}{H}P_I,_{X_I}P_J,_{X_J}\dot\phi_I\dot\phi_J\right)^\cdot.
\end{eqnarray}
This result is consistent with that obtained in multi-field inflation\cite{Langlois:2008mn, Langlois:2008wt, Langlois:2008qf}.
To distinguish the diagonal and off-diagonal parts of the matrices
above, we define
\begin{equation}
    \mathscr{N}_{IJ}=\mathcal{N}\delta_{IJ}+\mathcal{N}_{IJ},\;\;\;\;\mathscr{M}_{IJ}=\mathcal{M}\delta_{IJ}+\mathcal{M}_{IJ}.
\end{equation}
After assuming that all the slow-varying parameters defined in Sec. \ref{sect-intro} are small, we have
\begin{eqnarray}
  \mathcal{N} &=& 2\epsilon\frac{e_I^2}{\epsilon_I^2}\left[\epsilon\left(1+\frac{5}{6}\epsilon\frac{\eta_I}{\epsilon_I}\right)
  \left(1-\frac{\eta_I}{\epsilon_I}-\frac{h_I}{e_I}\right)\left(\frac{1}{\csi^2}+1\right)-\epsilon_I+\frac{\eta_I}{2}\right], \\
  \mathcal{N}_{IJ} &=& -H\left(\frac{1}{\csi^2}-1\right)e_Ie_J\left[1+\frac{1}{3}\epsilon\left(\frac{\eta_I}{\epsilon_I}+\frac{\eta_J}{\epsilon_J}
  \right)\right], \\
  \mathcal{M}_I &=& 3H^2\iota_I, \\
  \mathcal{M}_{IJ} &=&-3H^2e_Ie_J\left[
  1+\frac{1}{6}\sum_K\epsilon_K\left(\frac{1}{\csk^2}-1\right)\right].
\end{eqnarray}
Here, note that all these quantities are of order $\scro(\epsilon)$
if $\csi^2\sim\scro(1)$. In such a case, as we will see
below, the damping term has no correlations. For ordinary terms we have only preserved
the leading order to slow-variation parameters, but if a term
involves $\csi^{-2}$, we preserve one higher order. This is to
ensure the consistency when $\csi^2\sim\scro(1)$ fails, as in a
model with multiple DBI actions. Under the condition when
$\csi^2\sim\scro(\epsilon)$, all the terms should be preserved.

To investigate further we can derive the equations of motion for the
fields
\begin{eqnarray}\nn
  \ddot{\delta\phi}_I&+&\left\{\left[3H\left(1-\frac{2}{3}s_I\right)+(1-2\csi^2)\frac{\dot{P_I},_{X_I}}{P_I,_{X_I}}
  -2\csi^2\frac{P_I,_{X_IX_I}\dot{X_I}}{P_I,_{X_I}}\right]\delta_{IJ}\right.
  \\\nn
   &+&\left.\frac{\csi^2}{H}\sum_J\left(\frac{1}{\csj^2}-1\right)P_J,_{X_J}\dot\phi_I\dot\phi_J\right\}\dot{\delta\phi}_J\\\label{eom:general}
   &+&\left[\left(\frac{k^2\csi^2}{a^2}+\frac{\csi^2}{P_I,_{X_I}}\mathcal{M}\right)\delta_{IJ}
   +\frac{\csi^2}{P_I,_{X_I}}\sum_J\mathcal{M}_{IJ}\right]\delta\phi_J=0.
\end{eqnarray}
Note there is no summation over index ``$I$" even if it repeats in one
term. Here and throughout the paper Einstein's summation rule is not
applied for the index of fields (but still valid for repetition of
spatial sub/superscripts) and all the summations over capital Latin letters ``$I,J,...$" are
written explicitly.
In a generic case, we can rewrite the equations of motion as follows,
\begin{equation}\label{eom:unperturbed}
    \ddot{\delta\phi}_I+3H(1+\kappa_I)\dot{\delta\phi}_I+\left(\frac{k^2\csi^2}{a^2}+m_I\right)\delta\phi_I
    +H\sum_J\xi_{IJ}\dot{\delta\phi}_J+\sum_Jm_{IJ}\delta\phi_J=0.
\end{equation}
with
\begin{eqnarray}\label{mI}
m_I&=&\frac{\csi^2}{P_I,_{X_I}}\mathcal{M},\\\label{mIJ}
m_{IJ}&=&\frac{\csi^2}{P_I,_{X_I}}\sum_J\mathcal{M}_{IJ},\\\label{kappa}
    \kappa_I&=&-\frac{2}{3}s_I+\frac{1}{3}\eta_I-\frac{2}{3}\eta_I\csi^2-\frac{2\epsilon}{3}\left(1-\frac{\eta_I}{\epsilon_I}-\frac{h_I}{e_I}\right)
    (1-\csi^2),\\\label{xi}
    \xi_{IJ}&=&2\epsilon_I\frac{e_J}{e_I}\csi^2\left(\frac{1}{\csj^2}-1\right),
\end{eqnarray}
which are also small in inflation. We see there are two different
couplings in the damping and effective mass term respectively. Besides,
they both depend on time via the parameters. In principle, this kind
of equation is difficult to solve, if not impossible. Of course we
see that the coupling terms between different fields are all very
weak if we admire the slow-variation parameters to be small, which
may maintein the final solutions almost the same as those of $N$
independently evolving  fields. Further more, two special cases
which have specific limits of sounds of speed are most interesting.
One is that all fields $\phi_I$ have the canonical kinetic energy,
with $\csi\rightarrow1$. This is a decoupled multi-slow-roll model,
which is called assisted inflation by \cite{Liddle:1998jc} and
received much
focus\cite{Malik:1998gy,Kanti:1999vt,Copeland:1999cs,Green:1999vv}.
In such a case, $\xi_{IJ}\sim\scro(\epsilon^2)$ and therefore can be
totally neglected in the action, while $m_{IJ}\sim\scro(\epsilon)$.
We notice that only the correlations in mass term are preserved,
which are easy to manipulate. On the other hand, when all the fields
are DBI-type, as is discussed in \cite{Cai:2008if}, $\csi$ is very
small in the relativistic limit. For simplicity, we can suppose that
the sound speeds are as small as $\epsilon$, or even smaller. We
will shortly confirm that, the latter case is naturally suitable for analytical
calculation, while for former case one should impose further assumptions on the difference of $c_\mathrm{s}$'s. But whatever case it is, we see
$\xi_{IJ}\sim\scro(\epsilon)$ which is the main part of coupling,
and $m_{IJ}\leq\scro(\epsilon^{3/2})$ thus can be neglected.

One possible method is to consider the smallness of couplings, and
rotate the perturbative fields in field-space into a instantaneous
orthogonal basis which can generate the adiabatic perturbation which
is in the direction of the trajectory of fields in background, along
with $N-1$ entropy perturbations perpendicular to the direction of field trajectory.
After doing that, if we suppose the coupling terms
which are proportional to the derivatives of the rotating angles with respect to
time of the trajectory in field-space is small at the moment of
horizon crossing, we can neglect the coupling and treat the adiabatic
and entropy perturbations independently as free streaming quantum
fields. After some proper quantization, all the discussions are
similar to a $N$ decoupled multi-field
theory\cite{Langlois:2008mn, Langlois:2008wt, Langlois:2008qf}. Next
if we want to consider the effect of couplings that have been
omitted, one can treat such couplings as interacting vertices to the
quantum fields, and see how can the fields transfer to each other in
the quantum level\cite{Gao:2009qy}. Actually, if the fields have
different speeds of sound, it is difficult to do such rotations as
to make the adiabatic perturbation laid on the direction of
trajectory. This is because the angle we need to take our rotation
is different from that of the slope of trajectory, since sound
speeds will appear in the transformation and make the ``rotation'' anisotropic. Further
more, different sound speeds correspond to an anisotropic rescaling
in the rotation of basis. We will come back and explain this assert
in a two-field case by illustration later.

If we abandon the attempts to project the field into adiabatic
perturbation and entropy perturbations, we may find another route to
get our results. A reasonable method is to transform the
fields in order to eliminate the couplings, if possible, and then
treat the decoupled perturbations as free quantum fields, with a
final result of curvature perturbation $\mathcal{R}$ gained by
$\delta \mathcal{N}$ formalism. Before doing this, let us digress to take a
look at the solution if we discard all the couplings and see the
free fields limit of our theory. The result obtained here is also useful for future convenience. Under this situation, the equations
contains no coupling, while the only remaining task is to find the
solution of each independent field, and do quantization in
ultraviolet limit. We can consider the equations of motion as
\begin{equation}\label{eom:neglect couple}
    \ddot{\delta\phi}_I+3H(1+\kappa_I)\dot{\delta\phi}_I+\left(\frac{k^2\csi^2}{a^2}+m^2_I
    \right)\delta\phi_I=0,
\end{equation}
where the parameter $\kappa$ and $m$ have been defined in
(\ref{kappa}) and (\ref{mI}). This is a little different from the
standard form of single-field inflation because of an additional
term in the damping effect. We consider the fields moving in an
inflationary quasi-de Sitter phase, which means the comoving time
has a simple form\cite{Bartolo:2004if}
\begin{equation}\label{comoving time}
    \tau=-\frac{1}{Ha}\frac{1}{1-\epsilon}
\end{equation}
And we can postulating a rescaling on the fields,
\begin{eqnarray}\label{rescale}
    \chi_I(\tau)&=&z_I(\tau)^{1+\beta_I}\delta\phi_I,\\
    z_I&=&\frac{a\sqrt{P_I,_{X_I}}}{\csi^2}\approx\frac{ae_I}{\csi\sqrt{2\epsilon_I}},\\
\beta_I&=&\frac{3}{2}\kappa_I-h_I+s_I+\frac{\eta_I}{2}.
\end{eqnarray}
Note that, our rescaling contains an extra power of $\beta$ which is
a small quantity of order $\scro(\epsilon)$ comparing with the
standard definition of Mukhanov-Sasaki variable\cite{Mukhanov:1990me} due to the small shift in the
diagonal damping term of the Lagrangian. This extra power can be
determined by requiring the coefficients of $\chi_I'$ term in the
equation of motion vanish. The Mukhanov-Sasaki variable, as in the standard process before taking the quantization of the
fields, is actually defined via the normalizer before the kinetic terms
in the action(\ref{action:2nd order}). Therefore, we can convert the
equation into a standard form which is familiar to that in single
field theory:
\begin{equation}\label{chi background}
    \chi_I''+\left[k^2\csi^2-\frac{2}{\tau^2}\left(1+\frac{9}{4}\kappa_I+\frac{3}{2}\epsilon-\frac{m_I^2}{2H^2}\right)\right]\chi_I=0.
\end{equation}
This is just the Bessel equation of order
$\nu_I$ which will be determined below. The solution to this
equation is a linear combination of two independent Bessel
functions. However, to have the appropriate approximation behavior
which approaches the planar wave in Bunch-Davies vacuum with correct
normalization at early time $\tau\rightarrow-\infty$
\cite{Bunch:1978yq}, we need to choose the proper coefficients of
the Bessel functions, which corresponds to such a state in quantum
theory that it can minimize the energy density. One possible result
is to choose\cite{Mukhanov:1990me, Bartolo:2004if}
\begin{equation}\label{chi background solution}
\chi_I(\tau)=\frac{\sqrt{-\pi\tau}}{2}e^{i\frac{\pi}{2}\left(\nu_I+\frac{1}{2}\right)}H_{\nu_I}^{(1)}(-k\csi\tau),
\end{equation}
where $H_{\nu_I}^{(1)}$ is the Hankel function of the first type,
with
\begin{eqnarray}\label{nu}
\nu_I&=&\frac{3}{2}\sqrt{1+2\kappa_I+\frac{4}{3}\epsilon-\frac{8}{3}\csi^2\frac{\epsilon_I\iota_I}{e_I^2}}
\equiv\frac{3}{2}-\sigma_I,\\
\sigma_I&\approx&
s_I-\epsilon+\left(\csi^2-\frac{1}{2}\right)\eta_I+\epsilon\left(1-\frac{\eta_I}{\epsilon_I}-\frac{h_I}{e_I}\right)
    (1-\csi^2)+2\csi^2\frac{\epsilon_I\iota_I}{e_I^2}
\end{eqnarray}
as its order. Note that, $\sigma_I$ is a small quantity of order $\scro(\epsilon)$, which is irrelevant of the value of $\csi$ in a multi-DBI model when $\csi\sim\scro(\epsilon)$. Therefore, the amplitude of perturbation to the original field
$\phi_I$ is
\begin{equation}\label{amplitude delta phi}
    \mid\delta\phi_I\mid=2^{\nu_I-2}\frac{\Gamma(\nu_I)}{\Gamma(3/2)}
    \left(\frac{\csi\sqrt{2\epsilon_I}}{ae_I}\right)^{2(1+\beta_I)}\frac{(-k\csi\tau)^{\sigma_I}}{-\tau(k\csi)^{3/2}}
\end{equation}
Again we use the fact that both $\nu_I$ and $\kappa_I$ are very
small, thus we get the power spectrum of each field
\begin{eqnarray}\nn
    \mathcal{P}_{\delta\phi_I}&=&\left(\frac{H}{2\pi}\right)^2\frac{2}{\csi}\left[1+(\beta_I-2\sigma_I)\ln2-\sigma_I\psi\left(3/2\right)\right]
    \left(\frac{\epsilon_I}{e_I^2}\right)^{\beta_I}\left(\frac{H}{k}\right)^{2\beta_I}(1-\epsilon)^{2(1-\sigma_I)},\\\label{spectrum I background}
    &\approx&\left(\frac{H}{2\pi}\right)^2\frac{2}{\csi}\left(\frac{H}{k}\right)^{2\beta_I}\frac{\epsilon_I}{e_I^2}
\left[1-2\epsilon+\beta_I\ln2+\sigma_I\left(\gamma-2\right)+\beta_I\ln\frac{\epsilon_I}{e_I^2}\right],
\end{eqnarray}
where the second approximate equality holds at the moment when the
wavelength of the mode considered exits the sound horizon, i.e.
$k\csi=Ha$. $\psi$ is the digamma function and relates to
Euler-Mascheroni constant $\gamma$ as $\psi(3/2)=-\gamma+2-2\ln2$.
The dependence on the speed of sound is superficially different from
that of single-field DBI inflation. But when we take an extra
relationship $\dot{H}=-\sum_IX_I/\csi$ which holds in DBI-type
action, a new relation between the parameters $\csi$, $e_I$ and
$\epsilon_I$, holds as to leading order,
\begin{equation}\label{extra relation in DBI}
    2\frac{\epsilon_I}{e_I^2}=\csi.
\end{equation}
Therefore the dependence on $\csi$ in (\ref{spectrum I background}) is canceled, and
(\ref{spectrum I background}) will be coincident up to leading order with the result
in \cite{Chen:2006nt}. By virtue of $\delta \mathcal{N}$ formalism
\cite{Sasaki:1995aw, Lyth:2004gb}.
\begin{equation}\label{delta N}
    \zeta =\delta\mathcal{N}_e = \sum_I\mathcal{N}_e,_I\delta\phi_I +\frac{1}{2}\sum_{IJ}\mathcal{N}_e,_{IJ}\delta\phi_I\delta\phi_J+\cdots
\end{equation}
where $\mathcal{N}_e$ is the local e-folding number along a
trajectory from a spatially flat slice at a moment $t_\ast$ soon
after the relevant scale has passed outside the horizon during
inflation, to a uniform-density slice at another moment $t_c$ after
complete reheating when $\zeta$ has become a constant. So the power
spectrum curvature of perturbation $\zeta$ can be calculated by
\cite{Sasaki:1995aw}.
\begin{equation}\label{zeta general}
\mathcal{P}_\zeta=\left(\frac{H}{2\pi}\right)^2\sum_I\mathcal{N}_e,_I^2
\end{equation}
where $N_e,_I$ can be determined via
\begin{equation}\label{N,I general}
    H=-\sum_I\frac{\p \mathcal{N}_e}{\p\phi_I}\dot{\phi_I}=-2H\sum_I\frac{\p\mathcal{N}_e}{\p\phi_I}
    \frac{\epsilon_I}{e_I}\left(1-\frac{1}{3}\epsilon\frac{\eta_I}{\epsilon_I}
+\frac{2}{3}\epsilon\frac{h_I}{e_I}\right)
\end{equation}
in a general case.

Here for simplicity, we should consider the case when the $N$ fields
moves almost in the same manner. This is proved by
\cite{Liddle:1998jc} to be a late-time attractor for some
appropriate initial conditions. Later on we will demonstrate our
analytical calculation for a system with different sound speeds is
valid when $\epsilon_I-\epsilon_J\sim\scro(\epsilon^2)$, which
implies a field configuration like this. Under such assumptions, to
order $\scro(\epsilon)$ one have
\begin{equation}\label{N,I}
    \frac{\p\mathcal{N}_e}{\p\phi_I}=-\frac{1}{2N}\frac{e_I}{\epsilon_I}\left(1+\frac{1}{3}\epsilon\frac{\eta_I}{\epsilon_I}
-\frac{2}{3}\epsilon\frac{e_I}{h_I}\right)
\end{equation}
in a generic case. Therefore, together with (\ref{extra relation in
DBI}) we get
\begin{eqnarray}\nn
    \mathcal{P}^{(0)}_{\zeta}&=&\left(\frac{H}{2\pi}\right)^2\frac{1}{2N^2}\sum_I\frac{1}{\epsilon_I\csi}\left(\frac{H}{k}\right)^{2\beta}
    \\\label{P zeta uncoupled}&&\cdot\left[1+\frac{2}{3}\epsilon\frac{\eta_I}{\epsilon_I}
-\frac{4}{3}\epsilon\frac{e_I}{h_I}-2\epsilon+\beta_I\ln2+\sigma_I\left(\gamma-2\right)+\beta_I\ln\frac{\epsilon_I}{e_I^2}\right].
\end{eqnarray}
The superscript 0 denotes this power spectrum is the one for a
theory neglecting couplings. We can see if all the fields have the same
speed of sound, and $\epsilon_I\sim\epsilon/N$, (\ref{P zeta
uncoupled}) gives the results in multifield DBI
inflation\cite{Langlois:2008mn} which contains $N$ identifying
speeds of sound up to first order of slow variation parameters.

From (\ref{spectrum I background}), the scalar spectral index
$n_\mathrm{s}$ is
\begin{equation}\label{ns background}
n_{\mathrm{s}I}-1=-2\beta_I
=-2\epsilon\left(1-\frac{\eta_I}{\epsilon_I}-\frac{h_I}{e_I}\right)(1-\csi^2)+2\eta_I(1-\csi^2)
-2h_I.
\end{equation}
And the contribution to spectral index of $\zeta$ mainly comes from
the largest one of $n_{\mathrm{s}I}$'s, which is just the power
spectuum of curvature perturbation.


\subsection{Define New Fields}

To consider the coupling terms as a perturbations to equations of
motion (\ref{eom:unperturbed}), we could take a redefinition of inflaton
fields through a linear transformation which, up to leading order in
slow-variation parameters, can give a set of $N$ decoupled
equations of motion, with different coefficients appearing as new
speeds of sound and new effective mass terms. In general, this is
difficult to be realized. Fortunately, we are
dealing with a theory that is just perturbed from a decoupled one. As
the $N$ fields have tendency to move together, they have an even smaller
difference in the parameters. Intuitively the change in
eigenvalue should be small if the damping and mass matrix is not far
apart from a diagonal one. Actually Hoffman-Wielandt theorem in
matrix analysis tells us that the smallness of ``perturbation'' to a
diagonal matrix will cause small variation to the eigenvalues. To
make it clear let us denote $\lambda_I$ as the eigenvalues of a
diagonal matrix corresponding to a theory with no couplings, whose diagonal
entries are just the coefficients before each $\delta\phi$ term in
(\ref{eom:unperturbed}), and $\hat\lambda_I$ to be eigenvalues of
the ``perturbed'' matrices corresponds to the coupled coefficients
(\ref{eom:general}). Then, there exists a permutation $\sigma(I)$
for the new eigenvalues $\hat\lambda$ that satisfies\cite{Horn:1986}
\begin{equation}\label{Hoffman-Wielandt theorem}
    \left[\sum^N_{I=1}\mid\hat\lambda_{\sigma{I}}-\lambda_I\mid^2\right]^{1/2}\leq\parallel\delta
    M\parallel_2\sim\scro(N\epsilon).
\end{equation}
Here $\parallel\delta M\parallel_2$ is the Euclidean norm of a
perturbation matrix $\delta M$ which under our circumstance has the
magnitude of order $\scro(\epsilon)$. Thus, roughly speaking, the
variation in eigenvalue is a small quantity of order less than
$\scro(\epsilon)$, if $N$ is not very large. This fact ensures our
validity of our upcoming discussions on estimating the eigenvalues
after field redefinition.

Let us analyze the coupled equations under the fact that all the
coupling terms are small quantities of order $\scro(\epsilon)$. In
this case we expect the solutions will differ from the decoupled
ones very little. Later on we will see that, even under such case,
different sound speeds will bring a fruitful physical imprint. If
we impose a transformation in field space such that
\begin{equation}\label{field rotation}
    \delta\phi_I=\sum_mR^m_I\varphi_I,
\end{equation}
where $R$ is a non-singular transformation matrix. If we require the
transformation to be a representation of $SO(N)$ we can define a new
basis of perturbations which can be interpreted as adiabatic/entropy
directions, this matrix should be set orthogonal. But here it is not
possible in our case. Our purpose of transforming the fields is to decouple. And
unless for some special cases, the new fields has nothing to do with
adiabatic/entropy perturbations. Actually, if all the fields have
the same speed of sound, the fields redefinition (\ref{field
rotation}) could be an orthogonal transformation of $SO(N)$ group. And one
of the new basis would lay on the direction of tangent line of
trajectory in field space, which just duplicate the result in
\cite{Gordon:2000hv}. However, for fields with different speeds of
sound, we do not have such clear geometric explanation. In such a
case, we will see that the transformation is even only orthogonal up
to leading order $\scro(1)$. Later on we will see how we can explain
the new transformation in a field space.

To make our discussion clear we could require $R$ is also
slow-variant, which means $\dot{\mathbb{R}}\sim\scro(\epsilon)$.
This is reasonable in a quasi-de Sitter background, and its validity
will be confirmed after we get the concrete form of $\mathbb{R}$.
Thus, one can write the time derivative of new field transformed from $\mathbb{R}\delta\phi$ as
\begin{eqnarray}
  \dot{\delta\phi}_I &=& R^m_I\dot\vp_m+\dot R^m_I\vp_m, \\
  \ddot{\delta\phi}_I &\approx&R^m_I\ddot\vp_m+2\dot R^m_I\dot\vp_m.
\end{eqnarray}
Substitute into the equations of motion (\ref{eom:general}), we
have the equations for new variable $\vp$. Left multiply an inverse
transformation $(R^{-1})_n^I$ and take the summation over index $I$,
we get
\begin{eqnarray}\nn
  \ddot\vp_n &+& \sum_{m,I,J}(R^{-1})_n^I\left[\left(3H(1+\kappa_I)+2\dot R_I^m\right)\delta_{IJ}+H\xi_{IJ}R_J^m\right]\dot\vp_m
  \\\nn
   &+&\sum_{m,I,J}(R^{-1})_n^I\left[\left(\frac{k^2\csi^2}{a^2}R_J^m+m_I^2R_J^m+3H(1+\kappa_I)\dot
   R_J^m\right)\delta_{IJ}\right.\\
   &+&\left.m_{IJ}R_J^m+H\xi_{IJ}\dot R_J^m\right]\dot\vp_m=0.
\end{eqnarray}
If we need to decouple these equations into the form of
(\ref{eom:neglect couple}) and get a set of free propagating waves
in ultraviolet, we should find a way to diagonalize the two
coefficients in the square brackets before $\dot\vp$ and $\vp$ for
each equation at the same time. This is the simultaneous diagonalization of
two matrices, and is in principle impossible unless
the two matrices are commutative. The coefficient matrices does not
commute each other since $\xi_{IJ}$ is even asymmetric. All the
diagonal matrices are commutative. So the only remaining matrices
that may be non-commutative are those with $m_{IJ}$ and $\xi_{IJ}$
as their elements. Since their main part are diagonal elements that
surely commutes, we will justify that the complete matrices with a
little variation in off-diagonal parts are believed to
simultaneously diagonalizable up to $\scro(\epsilon)$. If we can
find a way to do this, then we can write
\begin{eqnarray}\label{diagonalize damping}
  \sum_I\left[2(R^{-1})_n^I\dot{R}_I^m+3H(R^{-1})_n^I\kappa_IR_I^m+H\sum_J(R^{-1})_n^I\xi_{IJ}R_J^m\right] &=&
  3HK_n\delta_n^m,\\\nn
  \sum_I\left[3H(R^{-1})_n^I(1+\kappa_I)\dot
  R^m_I+\frac{k^2}{a^2}(R^{-1})_n^I\csi^2R_J^m\right.&&
  \\\label{diagonalize mass}\left.+(R^{-1})_n^Im_I^2R_I^m+\sum_J(R^{-1})_n^Im_{IJ}^2R_J^m\right] &=&
  \left(\frac{k^2}{a^2}C_{\mathrm{s}n}+M_n\right)\delta_n^m.
\end{eqnarray}
Here we have neglect a $\xi\dot{\mathbb{R}}$ term since
$\dot{\mathbb{R}}$ is of order $\scro(\epsilon)$. Under such
transformation the equations of motion are completely decoupled,
\begin{equation}\label{eom:decoupled}
    \ddot\varphi_n+3H\left(1+K_n\right)\dot\varphi+\left(\frac{k^2}{a^2}C_{\mathrm{s}n}+M_n\right)\varphi_n=0.
\end{equation}
Now we use the Hoffman-Wielandt theorem which ensures us that the
solution to such eigenvalue equation of a perturbed matrix is a
small variation from the eigenvalue before. According to the
fundamental theorem of algebra, there are $N$ sets of such solutions
to (\ref{diagonalize damping}) and (\ref{diagonalize mass}), each
has a new speed of sound $C_{\mathrm{s}}$. The $\csn$'s are the
speeds of propagation of the new fields, akin to those of normal
modes in coupled oscillation system or phonon representation in
statistical system. To investigate whether the new representation
will change qualitatively the physics of an decoupled
system is an interesting question. For instance when $\csi$'s are
all close to 1, can we find a mode that has a very small speed of
sound $\csn$ that can possibly amplify the power spectrum and non-Gaussianity? We should
emphasize that although this is possible, it will not appear
in an weakly coupled system we are just dealing with since the reason
we've mentioned above. On the contrary, if the couplings are indeed
strong, the physics could be much more different. For example, in
hybrid inflation\cite{Linde:1993cn} where the two fields have
complicated coupling which is not small at the waterfall stage of
inflation, there are plentiful physical phenomena far beyond our
results here \cite{Lyth:2010ch, Lyth:2010zq, Gong:2010zf, Abolhasani:2010kn}.
Even in a weakly coupled system we are considering, analytical calculation is difficult
to do. But after we can diagonalize the damping and mass matrices
and take a set of new fields $\vp_n$ which can simplify the
equations of motion (\ref{eom:general}) to (\ref{eom:decoupled}),
the remaining task is very simple, just parallel to the discussion
under (\ref{eom:neglect couple}) since the form of the equation is
the same, only with a different definition of parameters $K_n$,
$C_{\mathrm{s}n}$ and $M_n$. The most difficult part of our method
is to find the transformation $\mathbb{R}$. In the next section we
will calculate in detail for an example of two-field inflation and
get a taste of such scheme.

\section{Two-Field Case}

To calculate the curvature perturbation at horizon-crossing in
detail we need the quantization of the fields in ultraviolet, which
requires a set of independent wave equations. As we mentioned above,
to simultaneously diagonalize so many matrices in (\ref{diagonalize
damping}) and (\ref{diagonalize mass}) seems to be an impossible
task. But after we notice all the off-diagonal entries are very
small and superadd an assumption that the diagonal entries are
almost the same up to $\scro(\epsilon)$, we can decouple the
equations in a natural manner.  In a two-field model, the fields
which drive inflation are $\phi_1$ and $\phi_2$, and a new definition
of fields provides two equations for the eigenvector of the
coefficient matrix, (\ref{diagonalize damping})(\ref{diagonalize
mass}). We have seen that for a generic value of $\csi$ the
off-diagonal damping and mass term will coexist. But for a model
with canonical kinetic energy, $\csi\rightarrow1$ and there is only
coupling in mass term\cite{Langlois:1999dw, Bartolo:2001rt, Byrnes:2006fr}. On the other hand if
$\csi$ is small as in the multi-DBI model, the coupling in damping
term survives. Let us work in the multi-DBI case, leaving aside the
matrix $m_{IJ}$ which has been proved to be up to
$\scro(\epsilon^{3/2})$, as has been discussed in the paragraph
under (\ref{mIJ}). Therefore the only remaining matrix to be
diagonalized is $\xi_{IJ}$, which has the eigenvalues
\begin{equation}\label{xi eigenequation}
    \left|\begin{array}{cc}
            \xi_{11}-\lambda & \xi_{12} \\
            \xi_{21} & \xi_{22}-\lambda
          \end{array}
    \right|=0.
\end{equation}
With the eigenvalues
\begin{equation}\label{xi eigenvalue}
    \lambda^{(1)}=2(\epsilon_1+\epsilon_2),\;\;\;\;\lambda^{(2)}=0.
\end{equation}
These eigenvalues corresponds to a transformation matrix
\begin{equation}\label{R}
    \mathbb{R}=\left(1+\frac{\epsilon_2}{\epsilon_1}\right)^{-1/2}\left(
        \begin{array}{cc}
          1 & -\frac{\displaystyle e_2c_{\mathrm{s}1}^2}{\displaystyle e_1c_{\mathrm{s}2}^2} \\
          \frac{\displaystyle\epsilon_2 e_1c_{\mathrm{s}2}^2}{\displaystyle\epsilon_1e_2c_{\mathrm{s}1}^2} & 1 \\
        \end{array}
      \right).
\end{equation}
Actually the eigenvector can only determine this matrix up to two normalization constants. Here we adopt a normalization such that it will preserve the determinant to be unity. This is not important in the quantization process, but will affect the calculation of power spectrum later. We know the similarity transformation $\mathbb{R}$ will diagonalize the matrix with elements $\xi_{IJ}$. Next we will show that it can also diagonalize the diagonal matrices with different diagonal entries, say
$\mathbb{K}=\mathrm{diag}(\kappa_1,\kappa_2)$, under some reasonable assumptions about the model. When the similarity matrix $\mathbb{R}$ operates on $\mathbb{K}$,
\begin{equation}
    \mathbb{R}^{-1}\mathbb{K}\mathbb{R}=\frac{1}{\epsilon_1+\epsilon_2}\left(
    \begin{array}{cc}
    \epsilon_1\kappa_1+\epsilon_2\kappa_2
    & \epsilon_1\frac{\displaystyle e_2c_{\mathrm{s}1}^2}{\displaystyle e_1c_{\mathrm{s}2}^2}(\kappa_2-\kappa_1) \\
    \epsilon_2\frac{\displaystyle e_1c_{\mathrm{s}2}^2}{\displaystyle e_2c_{\mathrm{s}1}^2}(\kappa_2-\kappa_1)
    &\epsilon_1\kappa_2+\epsilon_2\kappa_1 \\
    \end{array}
    \right).
\end{equation}
So a diagonal matrix which has different diagonal entries can remain
diagonal after the similarity transformation $\mathbb{R}$if
\begin{equation}\label{extra requirement on parameters}
\kappa_1-\kappa_2=\scro(\epsilon^2)
\end{equation}
with $\kappa_I$'s and $\epsilon_I$'s in the matrix are still of
order $\epsilon$. We will see, that for a specific multi-DBI
model, the assumption that $\kappa_1-\kappa_2$ as well as $s_1-s_2$
and $\eta_1-\eta_2$ are of order $\scro(\epsilon^2)$ is reasonable,
since the difference between two small positive quantities of order
$\scro(\epsilon)$ must be even smaller, if the fields have the
tendency to move together under an appropriate initial condition.
Besides, we will suppose that the differences between other
parameters like $m_1^2-m_2^2$ and
$c_{\mathrm{s}1}^2-c_{\mathrm{s}2}^2$, are also as small as
$\scro(\epsilon^2)$ for future convenience. Note, that this do not
implies $\epsilon_1-\epsilon_2\sim\scro(\epsilon^2)$ which is a much
more strict constraint on the evolution of fields. We will come back
to this issue and make some necessary justification to a specific two-field-DBI model later. At this moment, we just emphasize that this
assumption preserve the diagonal property of the diagonal matrices
in (\ref{diagonalize damping}) and (\ref{diagonalize mass})
involving $\kappa_I$, $\csi^2$ and $m_I^2$ automatically up to
$\scro(\epsilon^2)$ under a similarity transformation by any matrix
$\mathbb{R}$.

Next let us check whether this matrix can make $\mathbb{R}^{-1}\dot{\mathbb{R}}$ diagonal. The entries of the
matrix are
\begin{eqnarray}
  (\mathbb{R}^{-1}\dot{\mathbb{R}})_{11}
  &=&(\mathbb{R}^{-1}\dot{\mathbb{R}})_{22}=
  \frac{H\epsilon_2}{\epsilon_1+\epsilon_2}
  \left[(h_1-h_2)-\frac{1}{2}(\eta_1-\eta_2)-2(s_1-s_2)\right],\\
  (\mathbb{R}^{-1}\dot{\mathbb{R}})_{12} &=& \frac{H\epsilon_1}{\epsilon_1+\epsilon_2}
  \frac{e_2c_{\mathrm{s}1}^2}{e_1c_{\mathrm{s}2}^2}
  \left[(h_1-h_2)-2(s_1-s_2)\right],\\
  (\mathbb{R}^{-1}\dot{\mathbb{R}})_{21} &=&
  \frac{H\epsilon_2}{\epsilon_1+\epsilon_2}
  \frac{e_1c_{\mathrm{s}2}^2}{e_2c_{\mathrm{s}1}^2}
  \left[(h_1-h_2)-(\eta_1-\eta_2)-2(s_1-s_2)\right].
\end{eqnarray}
We see all the entries are proportional to the differences among some slow-variation parameters of the two fields, which are of order $\scro(\epsilon^2)$ as we mentioned before. Then we can assert that up to $\scro(\epsilon)$ the time derivative of $\mathbb{R}$ multiplied by its inverse in (\ref{diagonalize damping}) is negligible.

\begin{figure}
 \includegraphics[width=0.45\textwidth]{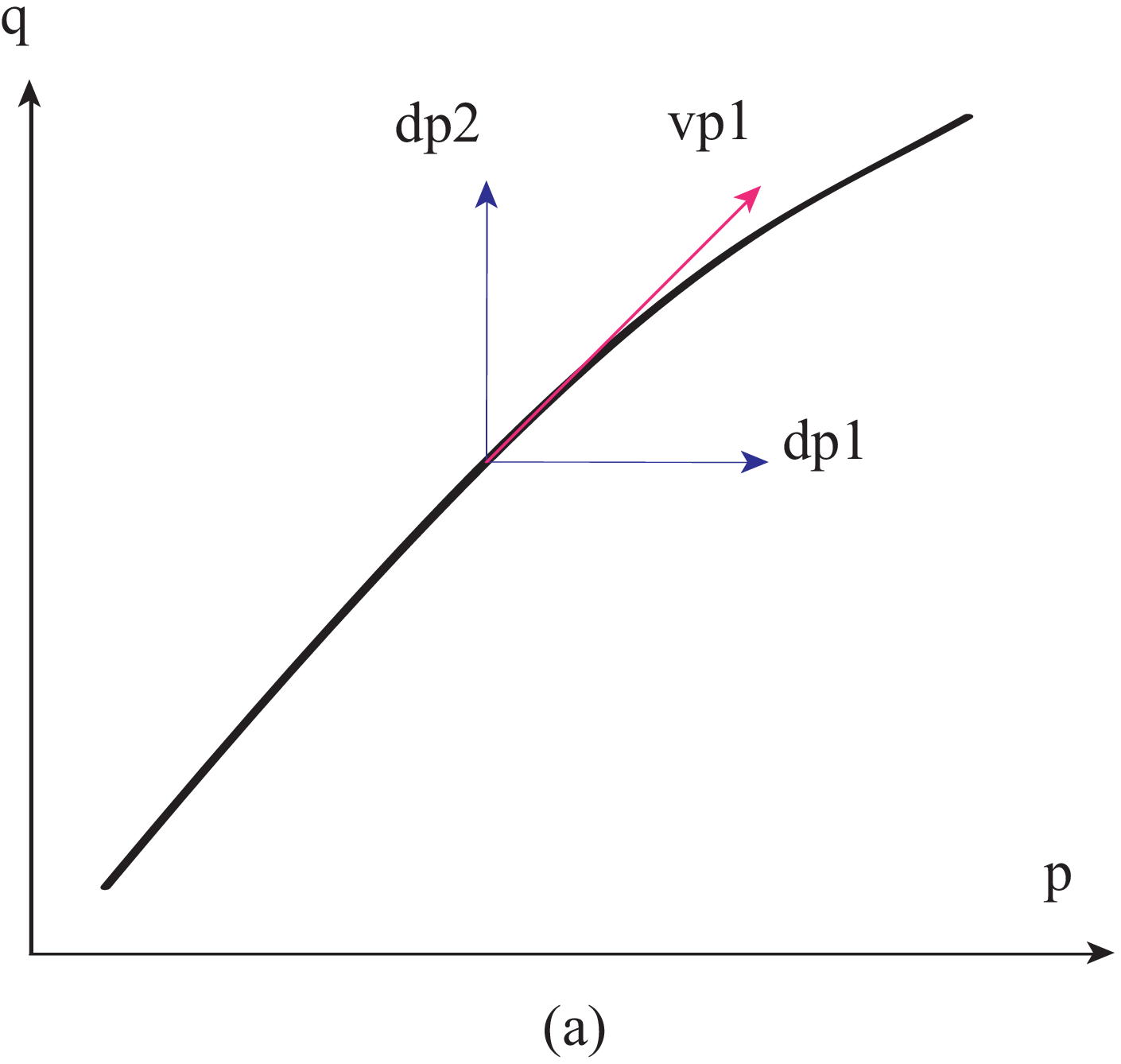}
 \includegraphics[width=0.45\textwidth]{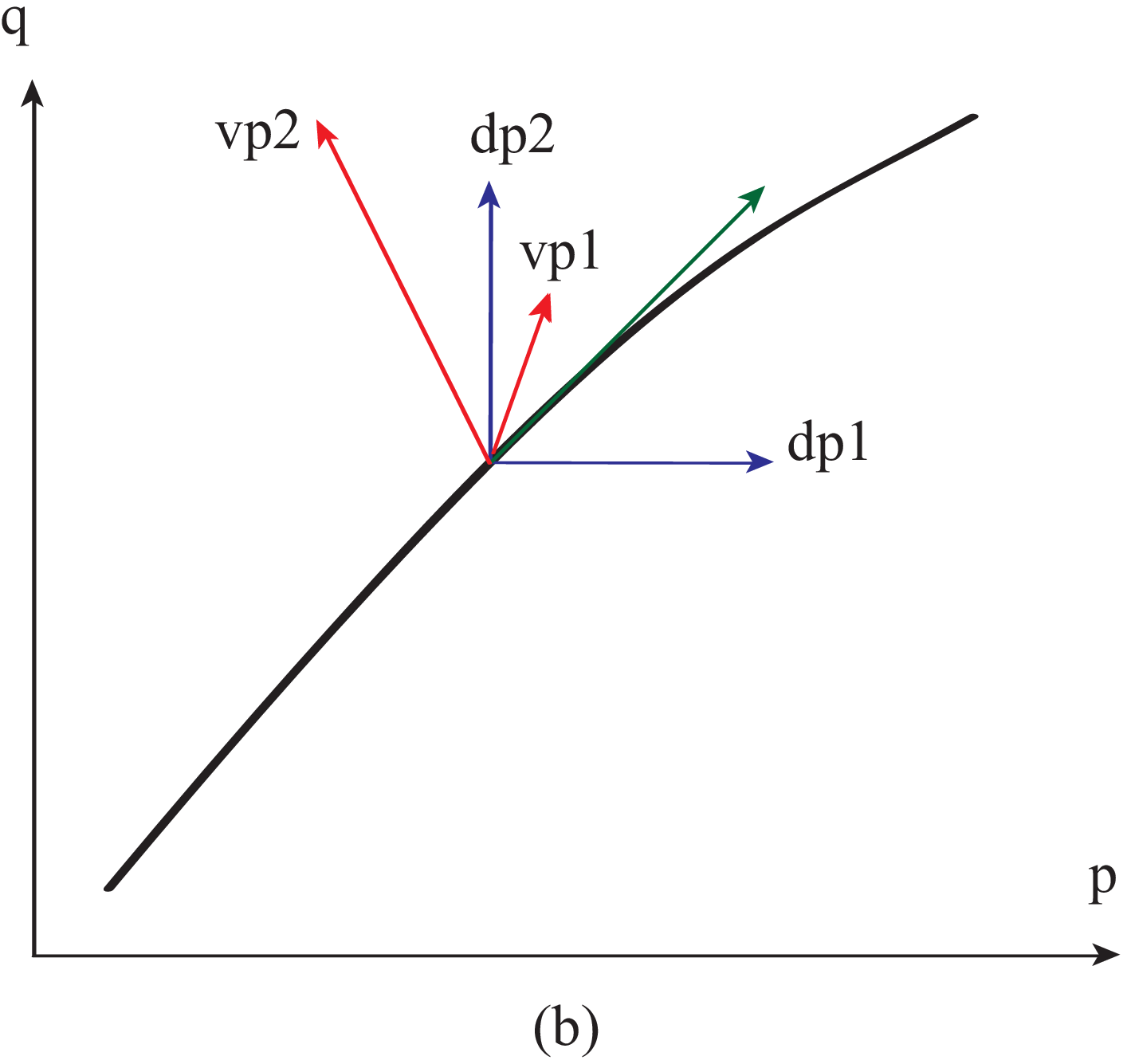}
  \\
  \caption{The transformation $\mathbb{R}$ is depicted by a transformation in field space of the perturbations.
  (a) denotes the trivial case when $\epsilon_1=\epsilon_2$, $e_1=e_2$ and $c_{\mathrm{s}1}=c_{\mathrm{s}2}$. We see for a totally
  symmetric configuration $\mathbb{R}$ poses an SO(2) rotation by
 an angle $\pi/4$, and $\phi_1$ is just the adiabatic perturbation
 in usual case. (b) shows a more interesting result when $\epsilon_1=(1/\sqrt{2})\epsilon_2$, $e_1=(1/\sqrt{2})e_2$ and
 $c_{\mathrm{s}1}=\sqrt{2}c_{\mathrm{s}2}$. Note this is to ensure $\dot\phi_1^2/\dot\phi_2^2=\epsilon_1c_{\mathrm{s}2}/(\epsilon_2c_{\mathrm{s}1})=1$ such that field are moving along a diagonal trajectory in fields space.
 We see that $\mathbb{R}$
 represents a transformation that rotates the field and rescales it at the same time with different rotating angle and anisotropic rescaling factor
 for two fields, i.e. the red arrow denoting the basis of new fields are not orthogonal, and the new fieds $\vp$ can not preserve the unitarity.}
\label{pert1}
\end{figure}

We will pause for a while and see the possible geometric explanation of transformation $\mathbb{R}$. We have seen that it is only orthogonal up to $\scro(\epsilon)$ under the assumption that $\epsilon_1-\epsilon_2\sim\scro(\epsilon^2)$. When $c_{\mathrm{s}1}=c_{\mathrm{s}2}$ as is the case of ordinary
assisted inflation, one have that the transformation $\mathbb{R}$ is just rotating the basis by $\pi/4$ and $\varphi_2$ is just the adiabatic perturbation with $\varphi_1$ the entropy perturbation which is negligible, as is depicted in Fig.\ref{pert1}(a). A more interesting case is the theory with different speeds of sound, as in (b).

After the transformation, one can find that two equations (\ref{eom:general}) are decoupled, which is just (\ref{eom:decoupled}),
\begin{equation}\nn
    \ddot\varphi_n+3H\left(1+K_n\right)\dot\varphi+\left(\frac{k^2}{a^2}C_{\mathrm{s}n}+M_n\right)\varphi_n=0
\end{equation}
with
\begin{eqnarray}\label{diagonal K}
  K_1 =
  \frac{\epsilon_1\kappa_1+\epsilon_2\kappa_2}{\epsilon_1+\epsilon_2}+\frac{2}{3}(\epsilon_1+\epsilon_2),&&
  K_2 = \frac{\epsilon_2\kappa_1+\epsilon_1\kappa_2}{\epsilon_1+\epsilon_2}, \\\label{diagonal Cs}
  C_{\mathrm{s}1}^2 = \frac{\epsilon_1c_{\mathrm{s}1}^2+\epsilon_2c_{\mathrm{s}2}^2}{\epsilon_1+\epsilon_2},
  &&
  C_{\mathrm{s}2}^2 =
  \frac{\epsilon_2c_{\mathrm{s}1}^2+\epsilon_1c_{\mathrm{s}2}^2}{\epsilon_1+\epsilon_2},\\\label{diagonal M}
  M_1=\frac{\epsilon_1m_1^2+\epsilon_2m_2}{\epsilon_1+\epsilon_2},&&
  M_2=\frac{\epsilon_2m_1^2+\epsilon_1m_2}{\epsilon_1+\epsilon_2}.
\end{eqnarray}
To go further let us study the change of the normalizer $z_{I}$ under the similarity transformation by $\mathbb{R}$. Note that, for example, when a normalizer matrix $\mathbb{Z}$, which is also diagonal, is multiplied by another matrix, say $\mathbb{X}$, the product is transformed under $\mathbb{R}$ by
\begin{equation}\label{normalizer}
    \mathbb{R}^{-1}\mathbb{ZXR}=\mathbb{R}^{-1}\mathbb{Z}\mathbb{RR}^{-1}\mathbb{XR},
\end{equation}
which is just multiplying each diagonal entries after the similarity transformation respectively. Denote by $Z_i$ the diagonal entries of $\mathbb{R}^{-1}\mathbb{Z}\mathbb{R}$,
\begin{eqnarray}\label{diagonal Z}
    Z_1(\tau)&=&\frac{a(\tau)}{\sqrt 2}\frac{c_{\mathrm{s}2}e_1\sqrt{\epsilon_1}+c_{\mathrm{s}1}e_2\sqrt{\epsilon_2}}
    {c_{\mathrm{s}1}c_{\mathrm{s}2}(\epsilon_1+\epsilon_2)},\\
    Z_2(\tau)&=&\frac{a(\tau)}{\sqrt 2}\frac{c_{\mathrm{s}1}e_2\epsilon_2\epsilon_1^{-1/2}+c_{\mathrm{s}2}e_1\epsilon_1\epsilon_2^{-1/2}}
    {c_{\mathrm{s}1}c_{\mathrm{s}2}(\epsilon_1+\epsilon_2)}.
\end{eqnarray}
After doing this, we can turn to the same process after equation (\ref{eom:decoupled}). Then we define new field variables
\begin{eqnarray}\label{rescale coupling}
    \chi_n(\tau)&=&Z_n(\tau)^{1+B_n}\varphi_n,\\
B_1&=&\frac{3}{2}K_1+\frac{c_{\mathrm{s}2}e_1\sqrt{\epsilon_1}\bigtriangleup_1+c_{\mathrm{s}1}e_2\sqrt{\epsilon_2}\bigtriangleup_2}
{c_{\mathrm{s}2}e_1\sqrt{\epsilon_1}+c_{\mathrm{s}1}e_2\sqrt{\epsilon_2}},\\
B_2&=&\frac{3}{2}K_2+\frac{c_{\mathrm{s}2}e_1\epsilon_1\epsilon_2^{-1/2}\bigtriangleup_1 +c_{\mathrm{s}1}e_2\epsilon_2\epsilon_1^{-1/2}\bigtriangleup_2
}{c_{\mathrm{s}1}e_2\epsilon_2\epsilon_1^{-1/2}+c_{\mathrm{s}2}e_1\epsilon_1\epsilon_2^{-1/2}},\\
\bigtriangleup_n&=&-\frac{1}{2}(\eta-\eta_n)+h_n-s_n,
\end{eqnarray}
 which satisfy the equations of motion
\begin{equation}\label{chi background}
    \chi_n''+\left[k^2C_{\mathrm{s}n}^2-\frac{2}{\tau^2}\left(1+\frac{9}{4}K_n+\frac{3}{2}\epsilon+\eta
    -\frac{1}{2}\bigtriangleup_n-\frac{M_n^2}{2H^2}+\frac{M_n^2}{H^2}(\epsilon-\eta)\right)\right]\chi_I=0.
\end{equation}
Now the equations have been decoupled. For each equation, the solution is a Hankel function with order $3/2-\varsigma_n$, where
\begin{equation}\label{theta}
    \varsigma_n=-K_n-\frac{2}{3}\epsilon-\frac{8}{9}\eta+\frac{2}{9}\frac{M_n^2}{H^2},
\end{equation}
Finally, we obtain the power spectra of the new fields $\varphi_i$ as follows,
\begin{eqnarray}\nn
  \mathcal{P}_{\varphi_1} &=& \left(\frac{H}{2\pi}\right)^2\frac{2}{C_{\mathrm{s}1}}\left(\frac{kC_{\mathrm{s}1}}{Ha}\right)^{2\varsigma_1}
  \frac{C_{\mathrm{s}2}^2(\epsilon_1+\epsilon_2)^2}{(C_{\mathrm{s}2}e_1\sqrt{\epsilon_1}+C_{\mathrm{s}1}e_2\sqrt{\epsilon_2})^2} \\
   &\cdot&\!\!\!\!\left[1-2\epsilon-2\varsigma_1\left(\ln2+\psi\left(3/2\right)\right)+\left(3K_1-2\bigtriangleup_1\right)
  \ln\frac{C_{\mathrm{s}1}C_{\mathrm{s}2}(\epsilon_1+\epsilon_2)}
  {C_{\mathrm{s}2}e_1\sqrt{\epsilon_1}+C_{\mathrm{s}1}e_2\sqrt{\epsilon_2}}\right],
  \\\nn
  \mathcal{P}_{\varphi_2} &=& \left(\frac{H}{2\pi}\right)^2\frac{2}{C_{\mathrm{s}2}}\left(\frac{kC_{\mathrm{s}2}}{Ha}\right)^{2\varsigma_2}
  \frac{C_{\mathrm{s}1}^2(\epsilon_1+\epsilon_2)^2}{(C_{\mathrm{s}1}e_2\epsilon_2\epsilon_1^{-1/2}
  +C_{\mathrm{s}2}e_1\epsilon_1\epsilon_2^{-1/2})^2} \\\nn
   &\cdot&\!\!\!\!\left[1-2\epsilon-2\varsigma_2\left(\ln2+\psi\left(3/2\right)\right)+\left(3K_2-2\bigtriangleup_2\right)
  \ln\frac{C_{\mathrm{s}1}C_{\mathrm{s}2}(\epsilon_1+\epsilon_2)}
  {C_{\mathrm{s}1}e_2\epsilon_2\epsilon_1^{-1/2}+C_{\mathrm{s}2}e_1\epsilon_1\epsilon_2^{-1/2}}\right].\\
\end{eqnarray}
The power spectra above are calculated at the moment when the wavelength exceeds the sound horizon, respectively. When one field, say $\varphi_1$ have exceeded the sound horizon, the power spectrum of $\varphi_1$ is frozen, and so the curvature perturbation calculated later should involve the value of $\mathcal{P}_{\varphi_1}$ calculated at $t_{\ast1}$ which satisfies $Ha(t_{\ast1})=kC_{\mathrm{s}1}$. The power spectrum of curvature perturbation should be calculated at the moment when the wavelength has exceeded all the sound horizons corresponding to different $\varphi$'s. The power spectra are nearly scale-invariant, with $-2\varsigma_n$ as their spectral index. The spectral index of curvature perturbation $\zeta$ is thus $-2\varsigma_n$ which has the largest absolute value.

\begin{figure}
  \includegraphics{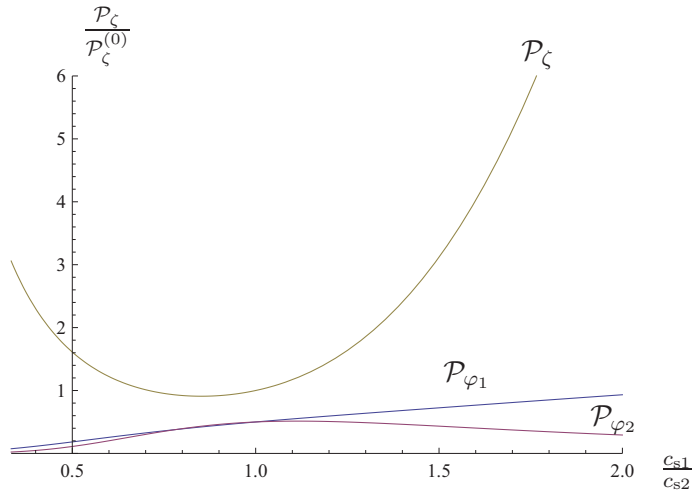}
  \caption{The dependence on $c_{\mathrm{s}}$ of the power spectrum. Both the power spectra of $\vp$ are normalized by $(H/2\pi)^2$. and $\mathcal{P}_\zeta$ is normalized by the un-coupled one
  (\ref{P zeta uncoupled}). There is severe divergence of $\mathcal{P}_\zeta$ when one of the sound speed vanish,
  but the only valid domain under our assumption is the case when $c_{\mathrm{s}1}\sim c_{\mathrm{s}2}$, i.e. in the neighborhood of $x\sim1$.
  We see the amplification by the reciprocal sound speed is much larger than that in a trivial $x=1$ case of ordinary unique-speed model. }\label{P-c}
\end{figure}

Using $\delta \mathcal{N}$ formalism, we can calculate the curvature perturbation after the wavelength being stretched out of the largest sound horizon. Note that to leading order
\begin{equation}
\zeta=\delta\mathcal{
N}_e=\sum_I\mathcal{N}_e,_I\delta\phi_I=\sum_I\mathcal{N}_e,_IR_I^m\varphi_m.
\end{equation}
There is no correlation between different $\vp$'s since they are independent kinetic modes as we defined. After using the additional relation (\ref{extra relation in DBI}) we can substitute $e_n$ in the expression by
$\sqrt{2\epsilon_n}/c_{\mathrm{s}n}$. Therefore we can simplify the final result
\begin{eqnarray}
  \mathcal{P}_\zeta &=& \left(\mathcal{N}_e,_1^2+\frac{\epsilon_2c_{\mathrm{s}2}^5}{\epsilon_1c_{\mathrm{s}1}^5}
  \mathcal{N}_e,_2^2\right)\mathcal{P}_{\varphi_1}
  +\left(\frac{\epsilon_2c_{\mathrm{s}1}^5}{\epsilon_1c_{\mathrm{s}2}^5}
  \mathcal{N}_e,_1^2+\mathcal{N}_e,_2^2\right)\mathcal{P}_{\varphi_2}.
\end{eqnarray}
The derivatives of the e-folding number can be calculated in a specific model in detail, for instance in a model with standard kinetic energy and separable potential\cite{Vernizzi:2006ve}. Since we only want to emphasize the effects caused by different sound speeds, again we use the assumption that both fields are moving in the same manner, i.e. in the diagonal trajectory in field space as in (b) of Fig.\ref{pert1}, and (\ref{N,I}) holds. This kind of motion is nearly adiabatic, which will always suppress the isocurvature perturbation inside the sound horizon. The final result is a little lengthy and instead of writing it here, we would rather focus on the dependence on different sound speeds by calculating $\mathcal{P}_\zeta/\mathcal{P}_\zeta^{(0)}$. By defining $c_{\mathrm{s}1}=xc_{\mathrm{s}2}$ one gets
\begin{equation}
    C_{\mathrm{s}1}^2\sim x,\;\;\;\;C_{\mathrm{s}2}^2\sim\frac{1+x^2}{1+x}.
\end{equation}
Now the power spectrum degenerates to
\begin{eqnarray}\label{P degenerated}\nn
\frac{\mathcal{P}_\zeta}{\mathcal{P}_\zeta^{(0)}}&=&
\frac{1}{2}\left(1+\frac{1}{x^4}\right)\frac{1}{\sqrt{x}} \left(\frac{1}{x^2}\sqrt{\frac{x}{1+x}}+\frac{x}{\sqrt{1+x^3}}\right)^{-2}\\&+&\frac{1}{2}\left(1+x^6\right)\sqrt{\frac{(1+x)^3}{1+x^3}} \left(x+\frac{1}{x^2} \sqrt{\frac{1+x^3}{1+x}}\right)^{-2},\label{P(x)}
\end{eqnarray}
which now just contains quantities up to leading order, and is in coincidence with (\ref{P zeta uncoupled}) when $x=1$, i.e. $c_{\mathrm{s}1}=c_{\mathrm{s}2}$. This normalized power spectrum seems to be divergent when $x\rightarrow0$ or $x\rightarrow\infty$. However, since we require $c_{\mathrm{s}1}^2-c_{\mathrm{s}2}^2\sim\scro(\epsilon^2)$,
the corresponding physical region is the neighborhood of $x\sim\scro(1)$. We depict the dependence on $x$ in Fig.\ref{P-c}.

We can see from Fig.\ref{P-c} that, there is a minimum of $\mathcal{P}_\zeta$ when $c_{\mathrm{s}1}=0.85c_{\mathrm{s}2}$, where $\mathcal{P}_\zeta=0.90\mathcal{P}_\zeta^{(0)}$. But the most interesting aspect is that if one of the speed of sound is a few times larger than the other (with themselves still as small as $\scro(\epsilon^2)$ so as to satisfy our condition required before), the power spectrum of curvature perturbation increases magnificently. We see from Fig.\ref{P-c} and (\ref{P(x)}) that if $x>1$, the contribution from $\vp_2$ exceeds that from $\vp_1$, and vis visa if $x<1$. Therefore we have encountered an interesting feature different from our intuition out of a theory without coupling, that the field with smaller speed of sound will not in general contribute more to the curvature perturbation. For instance, if $c_{\mathrm{s}1}>c_{\mathrm{s}2}$, it is the power spectrum from auxiliary field $\vp_2$ which contributes more to the final curvature perturbation, not $\delta\phi_1$.
 We also see that if $c_{\mathrm{s}1}/c_{\mathrm{s}2}$ deviates from 1, the power spectrum of curvature perturbation is amplified significantly, compared with the one without couplings. This just tells us that the effect given by the coupling terms may be large and can not be neglect in previous calculations. Since the power spectrum of curvature perturbation can be observed to rather high accuracy, this result actually tells us that we can loosen the constraints on the slow-variation parameters of individual inflaton even further.

\section{Conclusions and Discussion}

In the frame of inflationary cosmology involving multiple inflaton fields, it is possible to allow these fields possess their own sound speed parameters which characterize the propagation of field fluctuations correspondingly. It is important to study how these modes are related to the cosmological perturbation we may observe in CMB experiments. In the present paper we analyzed the dynamics of field fluctuations in a general scenario of MSI. Assuming no direct coupling terms among the inflaton fields, we noticed that these perturbation modes are generically coupled both through their effective mass terms and cosmic damping terms. Fortunately, we found that if these couplings are weak, the perturbation theory can be treated as a combination of normal modes, by which the curvature perturbation at the horizon-crossing can be calculated in detail up to leading order of slow-variation parameters via $\delta \mathcal{N}$ formalism for a specific model.

Specifically we studied a model of multi-speed inflation involving two DBI fields with their sound speeds being small and slow-varying. Our analysis showed that in the relativistic limit, the coupling between two field fluctuations is mainly contributed by the damping terms in their perturbation equations. We further studied the curvature perturbation in this model, and verified their primordial power spectra are nearly scale-invariant. This has interesting implications to the curvaton scenario\cite{Lyth:2001nq}, which suggests a light field in inflationary phase could seed entropy perturbation and convert it into curvature perturbation at the end of inflation\cite{Mollerach:1989hu, Linde:1996gt, Enqvist:2001zp, Lyth:2001nq, Moroi:2001ct}\footnote{Also see Ref. \cite{Cai:2011zx} for the extension of the curvaton scenario to non-inflationary cosmology.}. Namely, as shown in \cite{Li:2008fma, Zhang:2009gw, Cai:2010rt}, a light DBI field can realize the curvaton scenario and generate sizable non-Gaussianities of local and equilateral types after suitable fine-tuning on the model parameters. Our study also showed that, when the sound speed parameters for the inflaton fields are not equal, the perturbation modes would get frozen at different sound horizons. An important phenomenon is that the curvature perturbation of MSI could obtain an enhancement which depends on the ratio of the sound speeds.

The scenario of MSI is a very important branch of various models of multiple field inflation. In recent years, the inflationary models involving multiple field components have been studied extensively in the literature, namely, the analysis on the dynamics of N-flation \cite{Kim:2006ys}; multiple field inflation with particle decays \cite{Choi:2007fya}; the N-flation model in the frame of braneworld \cite{Panotopoulos:2007pg}; the model of staggered Nflation in the stringy landscape \cite{Battefeld:2008py}; dynamics of the model of coupled N-flation with canonical fields \cite{Ashoorioon:2008qr}; the inflation model constructed by a DBI field coupled to radiation \cite{Cai:2010wt}; and see \cite{Wands:2007bd, Langlois:2008sg} for comprehensive reviews on this field. It has been well realized that, for the model of multi-field inflation, when the sound speeds for the fields are the same, the field fluctuations are able to be rotated orthogonally to a basis in which the curvature and entropy fluctuations decouple explicitly, at least up to leading order. However, when we relax the scenario by allowing the sound speed parameters are different, our analysis implied that the rotation of the field fluctuations are no longer orthogonal but we still are able to find a transformation to decouple the fields under some conditions and obtain the curvature perturbation via $\delta\mathcal{N}$ formalism.

\begin{acknowledgements}
We are grateful to Antonio De Felice, Xian Gao and Tower Wang for useful discussions. We thank Yi-Fu Cai and Bin Chen for thoroughly reading the manuscript and providing valuable comments on our work. This work is supported by the NSFC Grant No.10975005 and RFDP. SP is supported by Scholarship Award for Excellent Doctoral Student granted by Ministry of Education of China.
\end{acknowledgements}

\end{document}